\documentclass[a4paper,11pt]{article}
\usepackage{amsfonts}
\usepackage{amsmath}

\input{tcilatex}

\begin{document}

\title{Uncertainty Relation in Flat Space from Holography}
\author{Jia-Zhong Chen$^{\thanks{%
E-mail: jzchen511@yahoo.com.cn}}$ and Withdraw \\
$^{\ast }$College of Chemistry and Chemical Engineering,\\
Northwest Normal University, \textit{Lanzhou 730070, P.R. China}\\
}
\maketitle
\date{}

\begin{abstract}
{\large Following the idea that information is the negative entropy, we
propose that the information and entropy of an isolated system can convert
into each other while the sum of them is an invariant for any physical
process. The holographic principle is then reformulated in the way that this
invariant is bounded by the Bekenstein-Hawking entropy} {\large of the
system. It is found that Heisenberg's uncertainty relation in quantum
mechanics can be derived from this bound.\ }
\end{abstract}

{\bfseries PACS number(s): }03.65.Ta,04.60.-m,03.67.-a

{\large \bigskip The concepts of information and entropy \cite{von
Neumann,Cover} play more and more important roles in physics both in
technical and theoretical aspects \cite{Landauer91}. It seems that
information is more fundamental than space as well as matter since each
physical law must be written in terms of a certain type of information such
as coordinate, curvature, energy, mass, etc.. Hence, in order to construct
the unified law of all interactions in nature, one should inquire what are
the informational contexts of space and matter and how to measure these
physical quantities in a unified framework of information. Ever since the
holographic principle was proposed \cite{tHooft93,Susskind94} the
information and entropy manifest themselves as key concepts as a guide for
constructing successful theory of quantum gravity (for recent review, see 
\cite{Bousso}). The information, known as how much uncertainty can be
eliminated in light of an observer, can be defined as the deviation of the
actual entropy of a system from the maximal entropy that the system may
contain \cite{Adami}. The feature of information is that an observer's
acquiring of information about an isolated system is always associated with
decreasing of uncertainty on the system. Since information describes the
observer's capability to predict the outcome of physical system some time
later, it is an observer-dependent quantity. However, the sum of the
information and entropy may be of physically significant if this sum is
independent of the observers. }

{\large In this Letter, we argue that information and entropy of a system
are two sides of one coin in the sense that they can convert into each other
in the process of measurement and evolution of the system while the sum of
them is an invariant. We show that this invariant is the maximal information
(or entropy) by which the holographic principle is reformulated. By applying
this assumption to a system of one particle, we derive uncertainty relation
from the holographic principle. }

{\large Let $p$ be the probability of obtaining the information $I$ for an
observer. The relation between them can be given by the fact that
probability is multipliable while information is additive: }

{\large 
\begin{equation}
I=k\ln (1/p).  \label{I1}
\end{equation}%
This is the well-known Shannon formula for information. By setting $k=1/\ln
2 $, one finds that Eq. (\ref{I1}) becomes 
\begin{equation}
I=\log _{2}(1/p).  \label{I2}
\end{equation}%
This implies $I=1$ when $p=1/2$, that is, the information to make decision
in two possible choices with equal probability is exactly $1$ bit. }

{\large Since the probabilities $p_{i}$ of acquiring the different
information $I_{i}$ vary or depend on the contexts of the information $%
I_{i}, $ $i=1,\cdot \cdot \cdot ,N$, the total amount of information should
be the direct sum of $I_{i}:I=\sum_{i=1}^{N}I_{i}$. The average information $%
\overline{I}$ is%
\begin{equation}
\overline{I}=\sum_{i=1}^{N}p_{i}I_{i}=-\sum_{i=1}^{N}p_{i}\log _{2}p_{i}.
\label{Iav}
\end{equation}%
It is evident that in the case of equal probabilities $p_{i}=p$ for all $%
p_{i}$%
\begin{equation}
\overline{I}=-\log _{2}p=-k\ln p=I_{i}.  \label{I5}
\end{equation}%
This enable us to look the $i$-th information $I_{i}$ as average
information. }

{\large We know that entropy is the measurement of how much a system is in
disorder while $I$ is exactly the opposite. For this reason, one can define
the information of an observer as the deviation of the actual entropy from
the maximal entropy \cite{Adami}. Clearly, this concept is well defined for
an isolated system only if we assume that the maximal entropy is fixed.
Motivated by the general consideration of the relationship between
information and entropy in physics and beyond, we propose an ansatz that the
sum of entropy and information (or, the total IE) is an invariant for an
isolated system:%
\begin{equation}
S+I=Invariant.  \label{I6}
\end{equation}%
in which the constant keep invariant during the system evolution. This
implies that for an observer the entropy $S$ of an isolated system will
decrease as information $I$ about it increases, and vice versa. In this
sense, information is nothing but IE though it may have not become into
knowledge about the system for a specified observer. In the following, we
will provide a demonstration of why this is so. }

{\large According to the standard thermodynamics, the entropy $S$ of an
isolated system is given by the Boltzman formula 
\begin{equation}
S=k_{B}\ln W,  \label{B7}
\end{equation}%
in which $W$ is the total number of the microscopic states corresponding to
a macroscopic state of the system. Notice that $p$ in Eq. (\ref{I1}) is the
probability of the macro-state appearing as an information, one has $%
p=W\cdot p_{0}$, where $p_{0}$\ is the probability of a micro-state
appearing. Hence, if we choose unified unit for information and entropy in (%
\ref{I1}) and (\ref{B7}) so that $k_{B}=k$, one finds%
\begin{equation}
k_{B}\ln W+k\ln (1/p)=k\ln (1/p_{0})=I_{0}.  \label{SI8}
\end{equation}%
For an isolated system with definite volume, $N$ is the total number of the
elementary particles, which is fixed. Since volume element $L$ in phase
space is finite and definite, $p_{0}=1/L^{N}$ is also definite. This shows
that Eq. (\ref{I6}) holds for the system. }

{\large We rewrite Eq. (\ref{SI8}) as 
\begin{equation}
S=I_{0}-I.  \label{S10}
\end{equation}%
Then when an observer finally acquires, by repeating measurements enough
times, maximal amount of information $I_{\max }$ about a given system, one
finds $I_{0}=I_{\max }$, that is 
\begin{equation}
S=I_{\max }-I.  \label{S11}
\end{equation}%
Eq. (\ref{S11}) indicates that entropy is the measurement of a part of
information that the observer is short of in trying to accurately describe
the states of system. }

{\large Now that the conservation equation (\ref{I6}) should hold for any
physical system on matter what underlying interactions involved, it must be
true for a black hole. For a black hole with horizon area $A$ we know from
the well-known Bekenstein-Hawking formula \cite%
{Bekenstein72,Bekenstein74,Hawking74} that its entropy is given by 
\begin{equation}
S_{BH}=\frac{k_{B}A}{4l_{p}^{2}},  \label{BH12}
\end{equation}%
in which $l_{p}=(G\hbar /c^{3})^{1/2}$ stands for the Planck length. In the
Planck unit this entropy is a quarter of the area of its horizon. Since we
know nothing about the black hole ($I_{BH}=0$), it follows from Eq. (\ref{I6}%
) that 
\begin{equation}
S_{BH}=I_{\max }.  \label{SI14}
\end{equation}%
This implies, together with Eq. (\ref{BH12}), that the information of an
outer observer about a black hole has completely converted into entropy
proportional to its horizon area. In other words, one can reinterpret Eq. (%
\ref{SI14}), by the assumption (\ref{I6}), as that the information has all
been saved on the horizon boundary of the black hole as its entropy. It is
clear that Eq. (\ref{BH12}) is the limit case of Eq. (\ref{I6}) in above
sense. Now that Eq. (\ref{BH12}) is believed to be valid without depending
the dynamical details of interactions \cite{Bousso,Wald99,Wald01}, one can,
in principle, assume that it applies for any system with strong
gravitational interaction, including the system of photons. }

{\large If one inputs energy from outer region into the system through its
boundary, the IE within the system will increase. Then, Eq. (\ref{SI14})
indicates that the total IE of an isolated system(or space-time region) with
boundary area $A$ is bounded up by the entropy $S_{BH}$ of a black hole with
same area of horizon, 
\begin{equation}
S+I\leq S_{BH}=\frac{k_{B}A}{4l_{p}^{2}}.  \label{SI16}
\end{equation}%
Clearly, this agrees with the holographic principle $S\leq A/4$ in Planck
unit since information $I$ of an observer about the system can not be
negative. The right side of Eq. (\ref{SI16}) is known as the holographic
bound \cite{Bousso}. We know that holographic principle provides for a
system an upper bound of information of 1 bit per Planck area $%
A_{p}=(2l_{p})^{2}$ on its boundary. We note here that Eq. (\ref{SI16}) has
reformulated the holographic principle by division the IE into two parts $S\ 
$and $I$ explicitly. }

{\large To see the implication of Eq. (\ref{SI16}), we reconsider\ the case
of black hole. From Eq. (\ref{S11}) we know that, in order to acquire
information from black hole, it is necessary for the outer observer to
reduce its entropy, or, equivalently, reduce the area of its horizon. This
can be exactly done by the process of Hawking radiation \cite{Hawking7475}.
Hence, in light of the formulation (\ref{SI16}), the Hawking process will
necessarily carry the information out of the black hole which has previously
disappeared in it. The black hole information paradox is then eliminated. }

{\large Let us consider, for simplicity, a free particle with mass M.
According to the holographic bound (\ref{SI16}) there exists a space-time
region $V$ enclosing this particle such that the IE of the particle would
not be more than that given by the area $A$ of the boundary $\partial V$ of $%
V$. Holographic principle implies that }$V${\large \ can not be a geometric
point since otherwise the IE of this particle will be equal or less that the
entropy bound }$0=A/4${\large \ , which contradicts with the fact that it
has the nonzero IE. In addition, a point-like particle can not be possible
since otherwise it appears as a singularity in space-time, which must be
surrounded by an even horizon (}$\partial V${\large )\ to avoid a bare
singularity. The radius of this region is the lower limit for position
observation. Therefore, one has following inequality for the entropy $S$ of
the particle 
\begin{equation}
S\leq \frac{k_{B}A}{4l_{p}^{2}}.  \label{S17}
\end{equation}%
}

{\large Even if we know nothing about the makeup of the particle we can
assume that the energy $E=Mc^{2}$ of the particle arises from the
contributions of its internal degrees of freedom constrained by the bound (%
\ref{S17}). We can then associate the energy $E$ with a temperature $T$
which corresponds to some unknown motion of these degrees of freedom. This
temperature $T$ should be in consistent with the standard relation of
entropy in thermodynamics: $S=\int_{0}^{E}dE/T$. Then, it follows from Eq. (%
\ref{S17}) that%
\begin{equation*}
\frac{Mc^{2}}{T}\leq \frac{k_{B}A}{4l_{p}^{2}},
\end{equation*}%
or, 
\begin{equation}
Mc^{2}\leq k_{B}T\cdot \frac{A}{4l_{p}^{2}}.  \label{Mc19}
\end{equation}
}

{\large For the internal motion of all degrees of freedom within
one-particle system, one can estimate that the energy of this system is
about }$Mc^{2}\ {\large \approx \ }${\large $k_{B}T$. Therefore, Eq. (\ref%
{Mc19}) becomes 
\begin{equation*}
\frac{A}{4l_{p}^{2}}\geq 1,
\end{equation*}%
or, using $l_{p}^{2}=G\hbar /c^{3}$, 
\begin{equation}
\frac{4\pi R^{2}}{4G\hbar /c^{3}}=\frac{\pi R^{2}c^{3}}{G\hbar }\geq 1,
\label{21}
\end{equation}%
in which $R$ is the characteristic radius of the region $V$. For an observer
who it is going to measure this particle $V$ is the hidden region and the
boundary $A=4\pi R^{2}$ is the horizon area of the particle with mass $M$.
Thus, $R$ should be given by Schwarzschild radius \cite{WaldGR}: $%
R=2MG/c^{2} $. If we rewrite Eq. (\ref{21}) as 
\begin{equation*}
\frac{Rc^{2}}{2G}\cdot \frac{2\pi cR}{\hbar }\geq 1.
\end{equation*}%
Then, one finds 
\begin{equation}
RMc\geq \frac{\hbar }{2\pi }.  \label{25}
\end{equation}%
}

{\large The fact that region $V$ is hidden for the observer outside implies
that the implication of the radius $R$ can be understood as the lower limit
of the position measurement implemented by the observer. Therefore, one can
identify $R$ as the position uncertainty $R=\Delta x$ of the particle with
respect to the observer. We also know \cite{WaldGR} that the hidden region $%
V $ given by Schwarzschild radius is the one that can not be measured even
by the fastest particles, namely, the photons. Therefore, $Mc=Mc^{2}/c$ is
the most-likely loss of the particle momentum during the measuring process
since the momentum transfer is maximal for the measurable particle with
vanishing rest mass. Hence, $Mc\approx \Delta p$. So, it follows from Eq. (%
\ref{25}) that 
\begin{equation}
\Delta x\cdot \Delta p\geq \frac{\hbar }{2\pi }.  \label{26}
\end{equation}%
This is the well-known uncertainty relation in quantum mechanics up to a
constant of order $1/\pi $. In spite of the presence of the mystical factor $%
1/\pi $ in the relation (\ref{26}) it is in consistent with the Heisenberg's
relation $\Delta x\Delta p\geq \hbar /2$. Since the validity of Heisenberg's
relation is mainly confined in the phenomena with the gravity relatively
weak in contrast with other interactions, one can look Eq. (\ref{26}) as the
uncertainty relation that may generally holds in quantum gravity. }

{\large In summary, we proposed an ansatz that the sum of information and
entropy of an isolated system is identically conserved in the process of
measurement and evolution of the system, by which the holographic principle
was reformulated in terms of the concept of the information entropy. It was
found that uncertainty principle between position and momentum can be
derived from the holographic principle. {Therefore, the uncertainty relation
in quantum theory is a manifestation of the finiteness of total information
that an observer can acquire by measuring a system. }\ }

\end{document}